\documentclass[prc,aps,preprint,showpacs,nofootinbib]{revtex4}
\usepackage{graphicx}

\begin{document}

\title{Selection Rule for Electromagnetic Transitions in Nuclear Chiral Geometry
}

\author{ Ikuko Hamamoto$^{1,2}$ }

\affiliation{
$^{1}$ {\it Division of Mathematical Physics, Lund Institute of Technology 
at the University of Lund, Lund, Sweden}   \\
$^{2}$ {\it The Niels Bohr Institute, Blegdamsvej 17, 
Copenhagen \O,
DK-2100, Denmark} \\ 
}


\begin{abstract}
In order to find the selection rules that can be applied to 
the electromagnetic transitions when the chiral geometry is achieved, 
a model for a special configuration in triaxial odd-odd nuclei is constructed 
which exhibits degenerate chiral bands with a sizable rotation.  A quantum
number obtained from the invariance of the Hamiltonian is given and the 
selection
rule for electromagnetic transition probabilities in chiral bands is derived 
in terms of this quantum number. Among the available candidates 
for chiral bands of odd-odd nuclei, in which the near degeneracy of two $\Delta
I = 1$ bands is observed, the measured electromagnetic properties 
of the two bands in 
$^{128}_{55}$Cs$_{73}$ and $^{126}_{55}$Cs$_{71}$ are consistent 
with the rules, while those of 
$^{134}_{59}$Pr$_{75}$ and $^{132}_{57}$La$_{75}$ are not. 
\end{abstract}


\maketitle

\section{Introduction}
The possible occurrence of chirality in nuclear structure was pointed out more
than ten years ago \cite{FM97}.  Since then, the observation of 
two almost degenerate $\Delta I$=1 bands possibly with the same parity 
has been reported, especially in the $A \approx 130$ odd-odd nuclei. 
Though the observed near degeneracy of the two bands is a primary indication of
chiral geometry, this geometry can be pinned down in a more definitive way if
electromagnetic transition probabilities expected for the chiral bands are
experimentally confirmed. 

Chirality in triaxial nuclei is characterized by the presence of 
three angular-momentum vectors, which are generally noncoplanar and 
thereby make it possible to define chirality. 
An example is shown in Fig. 1a. 
When chiral geometry is realized, observed two chiral-degenerate states 
are written as 
\begin{eqnarray}
\mid I + \rangle & = & \frac{1}{\sqrt{2}} 
\left( \mid IL \rangle + \mid IR \rangle \right) \nonumber \\
\mid I - \rangle & = & \frac{i}{\sqrt{2}} 
\left( \mid IL \rangle - \mid IR \rangle \right) 
\end{eqnarray}
where left- and right-handed geometry states are written as 
$\mid IL \rangle$ and $\mid IR \rangle$, respectively.  If there were  
no tunneling between
the $R$ and $L$ systems, the energies of the states associated with opposite
handedness would be degenerate and one obtains 
\begin{eqnarray*}
\langle IL \mid E2 \mid IR \rangle \approx 0 \quad\qquad \mbox{and} \quad\qquad 
\langle IL \mid M1 \mid IR \rangle \approx 0 
\end{eqnarray*}
for states with $I \gg 1$, where the electric-quadrupole and magnetic-dipole
operators are denoted by $E2$ and $M1$, respectively. 
If $\approx$ in the above expressions is replaced by =, 
two intra-bands or inter-bands
transitions or two static moments, which correspond to each other
within the observed pair of chiral bands, are equal.  The correspondence is 
illustrated in Fig. 1b.   
This is a trivial and straightforward rule in the case of the 
ideal chiral bands.   However, we want to take one step 
further with the selection rules.

\section{Model, Quantum Number and Selection Rule}
A limiting case of the particle-rotor model is considered which may be
applicable to the majority of odd-odd nuclei, in which the observation of the 
near degeneracy 
of two $\Delta I = 1$ bands has so far been reported\footnote{The contents 
of the present section are based on the work published in Ref. \cite{KSH04}.}.   
In our opinion an application of this
special theoretical limit outweighs a loss of generality. 

The model consists of a triaxially deformed core with $\gamma = 90^{\circ}$
coupled to one proton particle and one neutron hole in the same
single-$j$-shell.\footnote{The "rotation with $\gamma = 90^{\circ}$" is
equivalent to the "$\gamma = -30^{\circ}$ rotation" 
in the Lund convention except
that the intermediate axis is the quantization axis (taken to be the 3-axis).}
Taking the long, short, and intermediate axes of the triaxial body as the 1-, 2-
and 3-axes, the rotational Hamiltonian of the core is written as  
\begin{equation}
H_{rot} = \frac{\hbar^2}{8 \, \Im_{0}} [R^2_3 + 4(R^2_1 + R^2_2)]
\label{eq:hrot}
\end{equation}
where $\vec{R}$ expresses the core angular momentum and 
the $\gamma$-dependence of the hydrodynamical moments of inertia is
assumed.  The $\gamma$-dependence is approximately supported also 
by microscopic 
numerical calculations of moments of inertia.  In the case of a single-$j$-shell
configuration the triaxially quadrupole deformed potential can be written for
$\gamma = 90^{\circ}$ as \cite{HM83} 
\begin{equation} 
V^{\pi}_{sp} \propto (j^2_{p1} - j^2_{p2}) 
\label{eq:hjp}
\end{equation} 
for the proton particle and as 
\begin{equation} 
V^{\nu}_{sp} \propto (j^2_{n2} - j^2_{n1}) 
\label{eq:hjn}
\end{equation} 
for the neutron hole, using the fact that the one-particle matrix-elements of
($Y_{22} + Y_{2-2}$) are proportional to those of ($j^2_1 - j^2_2$).  
In Eqs. (\ref{eq:hjp}) and (\ref{eq:hjn}) $j_{p1}$ and $j_{p2}$ ($j_{n1}$ 
and $j_{n2}$) denote the components of the proton (neutron) angular-momentum
operator $\vec{j}_p$ ($\vec{j}_n$) along the 1- and 2- axes, respectively.  The
proportionality constants in (\ref{eq:hjp}) and (\ref{eq:hjn}), which are
linear in the quadrupole deformation parameter $\beta$, are positive and 
exactly the same if protons and neutrons are in the same single-$j$-shell.  
It is noted that in respective Hamiltonians of the proton particle, the neutron
hole and the core the energetically preferred
directions of relevant angular momenta are 
\begin{eqnarray*}
\vec{j}_p \, // \pm 2\mbox{-axis,} \quad\qquad \vec{j}_n \, 
// \pm 1\mbox{-axis,} 
\quad\qquad \mbox{and} \quad\qquad   \vec{R} \, // \pm 3\mbox{-axis.}
\end{eqnarray*}
However, one must look for 
the energy minimum for a given total angular momentum 
$I$ where $\vec{I} = \vec{R} + \vec{j}_p + \vec{j}_n$. 
Thus, the relative direction between the three vectors depends on the magnitude
of $I$.  (See Fig. 3.) 

The total particle-rotor Hamiltonian constructed from (\ref{eq:hrot}), 
(\ref{eq:hjp}) and (\ref{eq:hjn}) has the following properties. 

(i) $D_2$ symmetry \cite{BM75}, which leads to $R_3 = 0, \pm 2, \pm 4$, 
....... 

(ii) Invariance under the operation $A$ which consists of: 
(a) a rotation $exp(i(\pi /2)R_3)$ or $exp(i(3 \pi /2)R_3)$, combined with 
(b) an exchange of valence proton and neutron.

Denoting the operator that exchanges valence proton and valence neutron by $C$, 
we assign the values $C = + 1$ and $C = -1$ to the components of the intrinsic
neutron-proton wave functions which are symmetric and antisymmetric under
the operation $C$, respectively. 
Eigenstates of the total Hamiltonian have a quantum number $A = \pm 1$,
irrespective of whether the chiral geometry is achieved or not.  
The possible combinations of $R_3$ and $C$ for a given value of $A$
are shown in Table 1. 

For E2 transitions we take into account only the collective part, namely 
only the core contribution.  Then, in order to obtain non-zero B(E2) values, 
we must have both $\Delta C = 0$ and $\Delta R_3 \neq 0$.  The former is
required since the neutron and proton are spectators under the E2 transitions, 
while the latter is required since E2 matrix elements with $\Delta R_3 = 0$
vanish for the shape of $\gamma = 90^{\circ}$.  
Since the E2 operator can make only $|\Delta R_3| \leq 2$, 
the above condition of both $\Delta C = 0$ and $\Delta R_3 \neq 0$
leads to 
\begin{equation}
\Delta A \neq 0 \quad\qquad \mbox{in order to have} \qquad B(E2) \neq 0
\end{equation}
when we examine 
the contents of the eigenstates with a given $A$ value shown in 
Table 1.

The M1 transition operator in the particle-rotor model is written 
as \cite{BM75} 
\begin{equation}
\left( M1 \right)_{\mu} = \sqrt{\frac{3}{4 \pi}} \, \frac{e \hbar}{2mc} \left( 
(g_{\ell} -g_R) \, \ell_{\mu} + (g_s^{eff} - g_R) \, s_{\mu} \right)
\end{equation}
If we take, for example, $g_{\ell} = g_{\ell}^{free}$ and 
$g_s^{eff} = 0.6  \, g_s^{free}$ and $g_R = 0.5$, we obtain 
\begin{eqnarray*}
g_{\ell} - g_R = 0.5 \qquad & g_s^{eff} - g_R = 2.848 
\qquad & \mbox{for protons} \\
g_{\ell} - g_R = -0.5 \qquad & g_s^{eff} - g_R = -2.792 
\qquad & \mbox{for neutrons} 
\end{eqnarray*}
Then, we obtain $B(M1) \approx 0$  for $\Delta C = 0$, since the M1 operator is
almost antisymmetric under the exchange of valence proton and valence 
neutron.
Since the M1 operator can make only $|\Delta R_3| \leq 1$, we obtain 
\begin{equation}
[\, B(M1) \quad \mbox{with} \quad \Delta A \neq 0 \,] \quad \gg \quad 
[\, B(M1) \quad \mbox{with}
\quad \Delta A = 0 \,] 
\end{equation} 
examining the contents of eigenstates with a given $A$ value in Table 1.

The static quadrupole moment of a triaxially deformed core with $\gamma =
90^{\circ}$ vanishes. On the other hand, 
the static magnetic moment in the present model
is not negligible, since the magnetic moment operator can be written as 
\begin{equation}
\vec{\mu} \, = \, g_R \vec{R} + g_{\ell} \vec{\ell} + g_s \vec{s} \, = \, 
g_R \vec{I} + (g_{\ell} - g_R) \vec{\ell} + (g_s - g_R) \vec{s} 
\end{equation}
and has an extra term $g_R \vec{I}$ compared with the M1 transition operator 
in (6).

In the present model with the chiral geometry the exchange of the right-handed
with left-handed systems can be equally obtained by the exchange of the valence
proton and neutron while keeping the direction of $\vec{R}$ unchanged; see Fig.
1a. 
Since the rotation about the 3-axis, $exp((i(\pi / 2)R_3))$ or 
$exp((i(3 \pi / 2)R_3))$, does not affect chirality, 
the operation $A$ exchanges $|IR \rangle$ with $|IL \rangle$,
\begin{eqnarray*}
A|IL \rangle \propto |IR \rangle \qquad \mbox{and} \qquad A|IR \rangle \propto 
|IL \rangle 
\end{eqnarray*}
Then, the formation of chirality means that the two nearly degenerate states  
$|I+ \rangle$ and $|I- \rangle$  
have different eigenvalues of $A$
\begin{equation}
A|I+ \rangle = \pm \, |I+ \rangle \qquad \Leftrightarrow \qquad  
A|I- \rangle = \mp \, |I- \rangle
\end{equation}

Now, when bands are, by definition, arranged so that $\Delta I = 2$ E2
transitions are enhanced and always allowed within respective bands, 
the sign of $A$ in a given band must change at every increase of $I$ by 2, as
illustrated in Fig. 2a.  And, 
from the arguments described in the previous paragraphs 
the quantum number $A$ of the final state must have a
different sign from that of the initial state, in order to have 
strong $E2$ or $M1$ transitions.  
The consequence of the present selection rules with chiral geometry is
illustrated in Fig. 2a. 

We have chosen the particle-rotor Hamiltonian so that if at all possible
the chiral geometry may easily appear at moderate values of $I$.  
So, next, we try to numerically diagonalize 
our particle-rotor Hamiltonian, taking 
$j = h_{11/2}$ for both valence neutron and valence proton.  
In order to simulate a possible 
chiral nucleus
in the mass $A \approx 130$ region, the parameters $A$=130, $Z$=55, 
$\beta$=0.3, and $\Im_0$=8.55 $\hbar^2$/MeV are used.  
The results of the diagonalization are shown in Fig.2b. For $13 < I < 24$ the
approximate degeneracy of the two lowest 
bands, which are indicated by $f$ and $u$, 
is a good sign of the realization of chiral geometry.  
The bands in Fig. 2b are organized based on
energy at low spins close to the band-heads, while they are based on $B(E2; I
\rightarrow I-2)$ values over the degenerate spin range which is indicated by a
box. 
The characteristic features of electromagnetic transitions, 
which are expected for the chiral geometry as illustrated
in Fig. 2a, are produced exactly in the transitions between the 
states inside the box of Fig. 2b.

The relation between three angular momentum vectors,  
$\vec{R}$, $\vec{j_p}$ and $\vec{j_n}$, 
in the lowest-lying states for a given $I$ is illustrated in Fig. 3. 
At small rotation it is energetically cheaper for the vector $\vec{R}$ to be
placed on
the plane which is specified by $\vec{j}_p$ and $\vec{j}_n$, 
when a given value of $I$ has to be constructed.  See the left
figure of Fig. 3.  On the other hand, since 
at very high spins
the rotational energy becomes dominant, $\vec{R}$ points to the direction of
the 3-axis to save the energy, and both $\vec{j}_p$ and $\vec{j}_n$ start to
rotationally align also in the direction of the 3-axis. See the
right figure of Fig. 3.    
Consequently, only at moderate rotation the chiral geometry may be expected.  
This is what is seen in the calculated level scheme and transitions 
shown in Fig. 2b.  

It is noted 
that the invariance of the total Hamiltonian and the selection rule for
electromagnetic transitions described above also apply in the presence of pair
correlation that is treated in the BCS approximation.

\section{Comparison with Experimental Data and Perspectives}
There are a series of odd-odd nuclei in the $A \approx 130$ region \cite{KS01}, 
in which 
the energies of two $\Delta I = 1$ rotational bands are observed to be 
roughly degenerate (namely, up till a few hundreds keV energy difference), 
though in fact the equality of spin-parity has
hardly been experimentally proved in any of those nearly degenerate
states.
Up till several years ago the observed nearly degenerate pair-bands in 
$^{134}$Pr 
were indeed supposed to be the best example of chiral pair bands. 
However, the observed 
$B(E2;I \rightarrow I-2)$ values of two intra-band
transitions, which should be equal in the case of chiral pair bands, 
turned out to be
different at least by a factor of two \cite{DT07,PHS06}.   
Moreover, measured 
$M1$ transitions strongly violated the selection rule \cite{KSH04} described in
the present paper. Thus, the pair bands in $^{134}$Pr are no longer supposed to
be an example of chiral pair bands.  
However, if so, one may ask; if the observed bands in $^{134}$Pr are not 
the chiral pair bands, what are they ?  This is 
an even more interesting question.

Among the odd-odd nuclei in the $A \approx 130$ region 
there are at present two nuclei, 
$^{128}$Cs and $^{126}$Cs, in which 
not only the pair bands are nearly degenerate but also the measured 
electromagnetic transitions \cite{EG06,EG10} seem to be in agreement with the
present selection rule.  

Deviations of the actual situation in nuclei from the simple assumption made in
the present model may modify the selection rule described in the present paper. 
Nevertheless, the present selection rule should serve as a starting point for
the study of more complicated nuclear systems.  
In this connection the confirmation of chiral geometry in 
odd-A nuclei such as Nd isotopes, in which $\vec{j}_p$ in the present model is
replaced by the rotationally-aligned angular momentum of the S-band, 
is strongly wanted.  Thus, the
selection rule for electromagnetic transitions in that kind of chiral geometry
in odd-$A$ nuclei 
should be worked out
so that it can be used for examining available experimental data.

\newpage

\begin{table}[t]
\caption{Possible combinations of the $R_3$ and $C$ values for a given value of 
$A$, noting that the present particle-rotor Hamiltonian is invariant under the
operation $A$. }

\vspace{2pt}

\begin{tabular}{|c|c|c|} \hline

quantum-number $A$ &   $R_3$  & $C$ \\ \hline
+1 &  0, $\pm$4, $\pm$8, .....  & +1 \\
&  $\pm$2, $\pm$6, .....  & $-$1 \\ \hline 
-1 &  0, $\pm$4, $\pm$8, .....  & $-$1 \\
&  $\pm$2, $\pm$6, ..... &  +1 \\ \hline
\end{tabular}
\end{table}

\vspace{3cm}

\mbox{}

\newpage

\noindent
{\bf\large Figure captions}\\
\begin{description}
\item[{\rm Figure 1 :}]
(a) Sketch of the idealized chiral geometry.  The core, proton, 
and neutron angular momenta denoted by $\vec{R}$, $\vec{j}_p$ and $\vec{j}_n$,
respectively, are parallel to the principal axes of the triaxial body, 
which are labeled by the numbers inside parentheses. The right-handed system is
denoted by $R$, while the left-handed one by $L$.  (b) In the ideal chiral
geometry two transitions denoted by solid, dashed, dotted and dot-dashed lines, 
respectively, are equal for $I \gg 1$ and in the absence of tunneling between 
the $L$ and $R$ systems.
\end{description}

\begin{description}
\item[{\rm Figure 2 :}]
(a) Illustration of the selection rule for inter-bands and 
intra-bands $E2$ and $M1$ transitions in the chiral pair bands of 
the present model. The yrast band denoted by "y" and the yrare bands 
indicated by "ny" are slightly shifted in the figure, so as to simulate
practical examples.  Allowed $E2$ transitions
(with $\Delta I = 1$ and 2) and stronger M1 transitions (with $\Delta I = 1$)
are indicated by thick arrows.  For $\Delta I = 1$ transitions the allowed $E2$
transitions occur together with stronger $M1$ transitions, since both types of
transition need a change of $A$ quantum-number. 
Allowed but much weaker $\Delta I = 1$ $M1$
transitions are denoted by thinner arrows.  (b) Calculated level scheme and the
$E2$ and $M1$ transitions in the lowest two $\Delta I = 1$ 
bands, which are obtained from the diagonalization of our particle-rotor
Hamiltonian.  Arrows express transitions of strong $B(E2)$ or
$B(M1)$, while weaker $M1$ transitions in this numerical example are practically
forbidden transitions, and thus are not indicated in the figure.  
Solid and dashed lines represent levels with $A$=+1 and $-$1,
respectively, or vice versa. For E2 transitions, only the core contribution is
taken into account, while for M1 transitions $g_R$= 55/130 = 0.423 and
$g_s^{eff} = 0.6 \, g_s^{free}$ are used.  
\end{description}

\noindent
\begin{description}
\item[{\rm Figure 3 :}]
The relative direction of the three angular-momentum vectors in the
lowest-lying states for a given total angular momentum $I$, where 
$\vec{I} = \vec{R} + \vec{j}_p + \vec{j}_n$.  In the figure $\vec{j}_n$ and
$\vec{j}_p$ can be exchanged, for example.  In respective Hamiltonians of the
neutron hole, the proton particle and the core 
the energetically preferred directions are such that 
$\vec{j}_n$, $\vec{j}_p$ and $\vec{R}$ point to the $\pm$1 (long),
$\pm$2 (short), and $\pm$3 (intermediate) axes, respectively. 
\end{description}


\vspace{2cm}

\begin{thebibliography}{99}
\bibitem{FM97} S. Frauendorf and J. Meng, {\it Nucl. Phys.} {\bf A617} 
(1997) 131.

\bibitem{KSH04} T. Koike, K. Starosta and I. Hamamoto, {\it Phys. Rev. Lett.} 
{\bf 93} (2004) 172502.

\bibitem{HM83} I. Hamamoto and B. Mottelson, {\it Phys. Lett. B} {\bf 127} 
(1983) 281; {\bf 132} (1983) 7.

\bibitem{BM75} A. Bohr and B. R. Mottelson, {\it Nuclear Structure} (Benjamin,
Reading, Massachusetts, 1975), Vol.II.

\bibitem{DT07} F. Tonev {\it et al}., {\it Phys. Rev. C} {\bf 76} (2007) 
044313. 

\bibitem{PHS06} C. Petrache, G. B. Hagemann, I. Hamamoto and K. Starosta, 
{\it Phys. Rev. Lett.} {\bf 96} (2006) 112502. 

\bibitem{KS01} K. Starosta {\it et al}., {\it Phys. Rev. Lett.} {\bf 86} (2001) 
971; K. Starosta, a talk given in the present workshop and the references quoted
therein.  

\bibitem{EG06} E. Grodner {\it et al}., {\it Phys. Rev. Lett.} {\bf 97} (2006) 
172501. 

\bibitem{EG10} E. Grodner, a talk given in the present workshop 
and E. Grodner {\it et al}., to be published.   


\end{thebibliography}
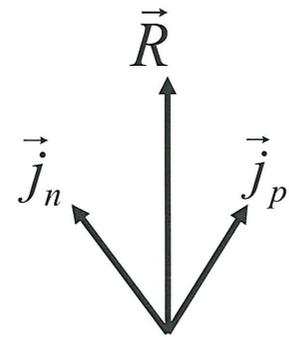
\end{document}